# Mining Reviews in Open Source Code for Developers Trail: A Process Mining Approach

**Patrick Mukala**
Departement of Computer Science, University of Pisa

**Article Info**

**ABSTRACT**

Audit trails are evidential indications of activities performers in any logs. Modern reactive systems such as transaction processing systems, management information systems, decision support systems and even executive management systems log activities of users as they perform their daily tasks for a number of reasons and perhaps one of the most important is security. In order to efficiently monitor and manage privacy and access to information, the trails as captured and recorded in these logs play a pivotal role in this regard. In Open Source realm, however, this is not the case. Although the objective with free software is to allow for access, free distribution and the rights to modify coding, having such audit trails can help to trace and understand how active members of these communities are and the type of activities they perform. In this paper, we propose using process mining to construct logs using as much data as can be found in open source repositories in order to produce a process model, also called a workflow net that graphical depicts the sequential occurrence of developers activities. Our method is exhibited through a simple algorithm called Act-Trace.

.

*Corresponding Author:*

Patrick Mukala,
Departement of Computer Science,
University of Pisa,
Largo Pontecorvo 5, 56127 Pisa PI
Email:patrick.mukala@gmail.com, mukala@di.unipi.it

## 1. INTRODUCTION

Process Models as graphical models present a viable means of visualizing traces of learning processes' activities. Making use of data recorded for OSS projects, process mining allows producing some representation sketches about the inherent activities. However, to date, there have been very limited attempts in process mining these repositories. A number of challenges or constraints could explain this. Perhaps one of the notable factors is the structure of data records in OSS repositories. The figure below illustrates the basic building blocks of process mining [9].

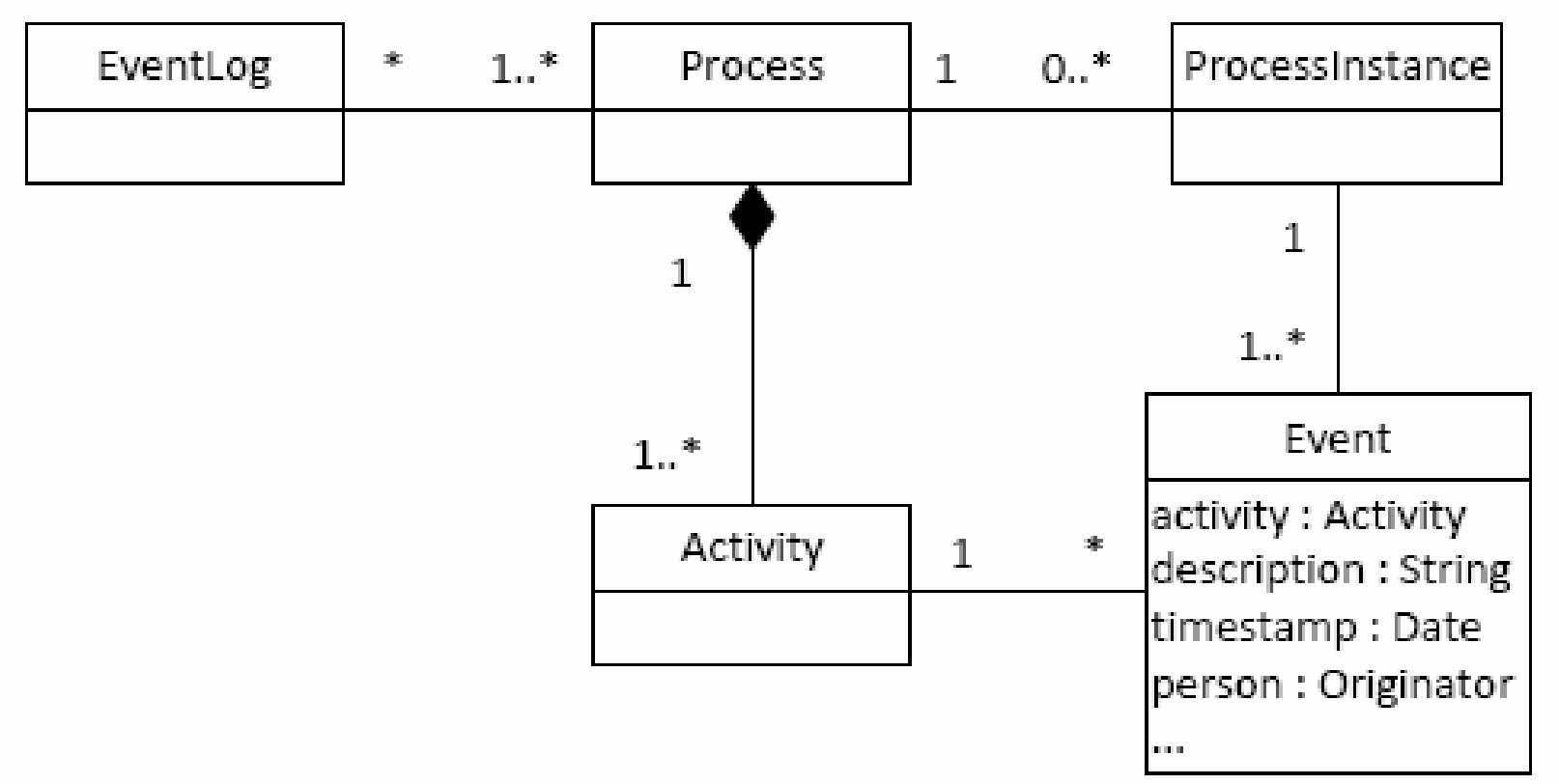

Figure 1: Process Mining Meta Data Model

The idea as expressed in the model is that a log, an event log that is ready for process mining should abide by a number of structural properties to facilitate its processing and analysis thereof by the existing tools. Simply put, an event log should contain data organized and clustered in processes, each of these processes have instances uniquely identifiable with a set of activities. A process instance can also be referred to as a case instance includes a number of events that consist of activities being executed at a given point in time. An example could be a log of an insurance company might contain information about a billing and refund process. A refund process has a number of process instances uniquely identified by the claim number. Activities that should be executed in the refund process may include registering the claim, and checking the insurance policy. An example of an event is "On Thursday September 23, 2010 Alice checks the insurance policy of the persons involved in claim 478-12" [9].

Given such information, the goal of process mining can be, through its techniques, to derive abstract representations of the process control flow, detect relations between the individuals involved in the process and their tasks, and infer data dependencies between different process activities.The chief objective in this paper is to exemplify the adoption of process mining techniques for the analysis of OPEN Source Files. While this area can store volumes of data containing patterns of online collaboration, limited approaches exist that aim to provide evidence-like activities trace between participants.

We propose a simple algorithm called Act-Trace that demonstrates steps taken to gather data needed for our case study in order to generate an event log needed for analysis in Disco process mining tool in [10]. Although the aim is to demonstrate how Act-Trace works, we also discuss a number heuristics and ways forward as pointers for possible logs that can be constructed in OSS data.

The rest of the paper is structured as follows: in section 2 we give an overview of process mining, section 3 introduces the assumptions and heuristics for log construction in Open Source Software, we also introduce the algorithm and give the results of analysis as obtained, and we conclude in section 4.

## 2. PROCESS MINING FOR ACTIVITY TRACING

Process mining is used as a method of reconstructing processes as executed from event [4]. These logs can be generated from process-aware information systems such as Enterprise Resource Planning (ERP), Workflow Management (WFM), Customer Relationship Management (CRM), Supply Chain Management (SCM), and Product Data Management (PDM) [3]. The logs contain records of events such as activities being executed or messages being exchanged on which process mining techniques can be applied in order to discover, analyze, diagnose and improve processes, organizational, social and data structures [5].

In [3], the goal of process mining is described to be the extraction of information on the process from event logs using a family of a-posteriori analysis techniques. These techniques enable the identification of sequentially recorded events where each event refers to an activity and is related to a particular case (i.e., a process instance). They also can help identify the performer or originator of the event (i.e., the person/resource executing or initiating the activity), the timestamp of the event, or data elements recorded with the event.

Current process mining techniques evolved from the work done in [4] where the purpose was to generate a workflow design from recorded information on workflow processes as they take place. Assuming that from event logs, each event refers to a task (a well-defined step in the workflow), each task refers to a case (a workflow instance), and these events are recorded in a certain order. The work in [4] combines the techniques from machine learning and Workflow nets in order to construct Petri nets that provide a graphical but formal language for modeling concurrency as seen in the figure below.

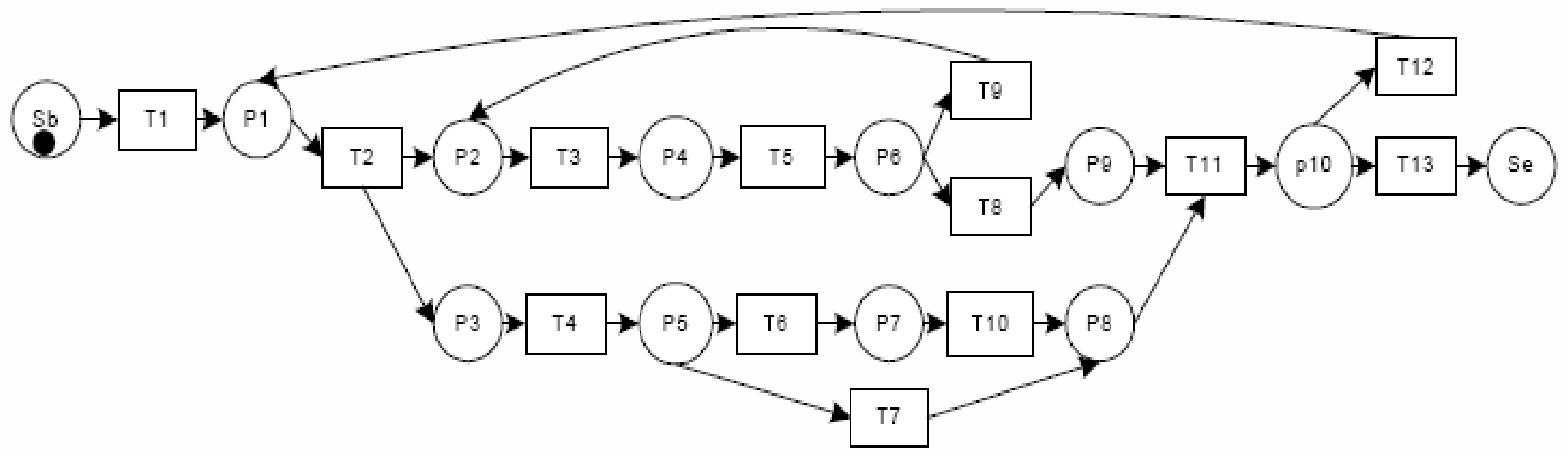

**Figure 2: example of a workflow process modeled as a Petri net**

The preliminaries of process mining can be explained starting with the α-algorithm of which formalization is given below.

Let W be a workflow log over T. α(W) is defined as follows.

1. $T_W = \{ t \in T \mid \exists_{\sigma \in W}\, t \in \sigma\}$,
2. $T_I = \{ t \in T \mid \exists_{\sigma \in W}\, t = \mathit{first}(\sigma) \}$,
3. $T_O = \{ t \in T \mid \exists_{\sigma \in W}\, t = \mathit{last}(\sigma) \}$,
4. $X_W = \{ (A,B) \mid A \subseteq T_W \wedge B \subseteq T_W \wedge \forall_{a \in A} \forall_{b \in B}\, a \rightarrow_W b \wedge \forall_{a1,a2 \in A}\, a_1 \#_W a_2 \wedge \forall_{b1,b2 \in B}\, b_1 \#_W b_2 \}$,
5. $Y_W = \{ (A,B) \in X \mid \forall_{(A',B') \in X} A \subseteq A' \wedge B \subseteq B' \Rightarrow (A,B) = (A',B') \}$,
6. $P_W = \{ p_{(A,B)} \mid (A,B) \in Y_W \} \cup \{i_W, o_W\}$,
7. $F_W = \{ (a, p_{(A,B)}) \mid (A,B) \in Y_W \wedge a \in A \} \cup \{ (p_{(A,B)}, b) \mid (A,B) \in Y_W \wedge b \in B \} \cup \{ (i_W, t) \mid t \in T_I\} \cup \{ (t, o_W) \mid t \in T_O\}$, and
8. $\alpha(W) = (P_W, T_W, F_W)$.

The sequence of execution of the α-algorithm goes as follows [6]: the log traces are examined and in the first step, the algorithm creates the set of transitions *($T_W$)* in the workflow, (Step 2) the set of output transitions ($T_I$) of the source place, and (Step 3) the set of the input transitions ($T_O$) of the sink place. In steps 4 and 5, the α-algorithm creates sets ($X_W$ and $Y_W$, respectively) used to define the places of the mined workflow net. In Step 4, it discovers which transitions are causally related. Thus, for each tuple (A, B) in $X_W$, each transition in set A causally relates to all transitions in set B, and no transitions within A (or B) follow each other in some firing sequence. Note that the OR-split/join requires the fusion of places. In Step 5, the α-algorithm refines set XW by taking only the largest elements with respect to set inclusion. In fact, Step 5 establishes the exact amount of places the mined net has (excluding the source place iW and the sink place oW. The places are created in Step 6 and connected to their respective input/output transitions in Step 7. The mined workflow net is returned in Step 8 [6].

From a workflow log, four important relations are derived upon which the algorithm is based. These are $>_W$, $\rightarrow_W$, $\#_W$, and $||_W$ [6]. In order to construct a model such as the one in Figure 7 on the basis of a workflow log, the latter has to be analyzed for causal dependencies [7]. For this purpose, the Log-based ordering relations notation is introduced:

Let W be a workflow log over T, i.e., $W \in P(T*)$. Let $a, b \in T$:

- a $>_W$ b if and only if there is a trace $\sigma = t_1 t_2 t_3 \ldots t_{n-1}$ and $i \in \{1,\ldots,n-2\}$ such that $\sigma \in W$ and $ti = a$ and $ti+1 = b$,
- a $\rightarrow_W$ b if and only if a $>_W$ b and b $>_W$ a,
- a $\#_W$ b if and only if a $>_W$ b and b $>_W$ a, and
- a $||_W$ b if and only if a $>_W$ b and b $>_W$ a.

Considering the workflow log W = {ABCD, ACBD, AED}, relation $>_W$ describes which tasks appeared in sequence (one directly following the other). Clearly, A $>_W$ B, A $>_W$ C, A $>_W$ E, B $>_W$ C, B $>_W$ D, C $>_W$ B, C $>_W$ D, and E $>_W$ D. Relation $\rightarrow_W$ can be computed from $>_W$ and is referred to as the (direct) causal relation derived from workflow log W. A $\rightarrow_W$ B, A $\rightarrow_W$ C, A $\rightarrow_W$ E, B $\rightarrow_W$ D, C $\rightarrow_W$ D, and E $\rightarrow_W$ D. Note that B $\rightarrow_W$ C because C $>_W$ B. Relation $_W$ suggests potential parallelism.

## 3. PROCESS MINING OPEN SOURCE CODE REVIEWS

The primary task in process mining OSS repositories is to generate an event log. We can note assert with certainty that in OSS data records files adhere to the basic structure as required in the Process Mining Meta Data Model in the figure above. A number of challenges in this regard include the disparity and sometimes the incompleteness of data [sometimes when only MailingLists or Code Reviews or Commits or CVS comments are provided separately without any link to other repositories] or its lack thereof, the identification and specification of processes or cases or even events in these repositories.

However, all these challenges can be overcome depending on the purpose of process mining with regard to the type of analysis one wishes to conduct and the goals to be reached. Therefore, as long as the objectives are clear, we can introduce a number of guidelines that will help in constructing process models from OSS data:

→ Process instances, in our case, could be commits made by a contributor for a given period of time. Commits can be regarded as any software artifacts, files or piece of code submitted by any project participant as part of the development cycle.

→ Alternatively, Process instances could group email messages about a Topic or comments in a discussion forum about a specific question for a given period of time. Our analysis could detect and identify activities we can cluster as specified by the requirements we try to identify from the data.

→ An OSS project, can be a process where we cluster, all activities that have taken place since its inception and try to group them into sub processes as needed.

→ Even further, commits about a component or a bug is material for a process. As with previous propositions, the goal is to cluster the traces of activities and graphically display them through a model called a process model.

All of these assumptions can be verified and evaluated accordingly with data. More precisely and within the confine of our research objectives:

→ These assumptions provide a way to construct process models and identify at which phase of the learning process identified activities can be assimilated to.

### 3.1. Heuristics

Mining software repositories is achieved almost entirely based on a number of heuristics. An extensive review of current approaches to mining data in OSS is based on some initial assumptions of which some have been verified and some not really verified. Given the nature of OSS data available today, we can capture as much as possible the traceability of the review activities and validate the models as needed. Nevertheless, our approach should be able to capture and explain the patterns as needed based on the Heuristics as summarized in the table below.

A number of assumptions based on the literature and heuristics guide the generation of event logs. Specifically, with no automatic analysis of OSS repositories and given the availability of data, the logs should reflect the following activities as summarized in the table below

**Table 1: Process Mining OSS repositories: Heuristics**

| LEVEL OF TRACEABILITY ⇒ REPOSITORY | TYPE OF ACTIVITIES | PROCESS/EVENT CASE |
|---|---|---|
| ***Initiation*** ⇑ Messages (Mailing/Forums/CVS comments) | Observe/Make Contact : INITIATOR (read, comment…), RESPONDER (read, comment, post…) | Message Title, Discussion Topic, Artifact Thread |
| ***Progression*** ⇑ Messages (Mailing/Forums/CVS comments) Internet Relays Messages, Bug Tickets, CVS/SVN | **Reply/Post** : INITIATOR(send, reply, post), RESPONDER(reply, post, report). **Apply :** INITIATOR (run, analyze, comment…), RESPONDER (run, analyze, comment…) | Commit, Message Title, Discussion Topic, Artifact Thread, |
| ***Maturation*** ⇑ MailingLists/Forums/CVS comments/ Internet Relays Messages, Bug Tickets, CVS/SVN | **Analyze/Review/Revert** : INITIATOR (review ,modify, submit...), RESPONDER (Revert, analyze, guide…). **Develop/Commit** : INITIATOR(modify, report, develop), RESPONDER(run, analyze, report) | Message Title, Discussion Topic, Artifact Thread, Internet Relays Messages |

This table summarizes in a general way some assumptions on the types of activities that can be traced, where they can possibly be retrieved and the possible candidate for event case as will be expressed in the process models. The idea is that activities expressing the initial phase of the process can be traced by looking at Messages and remarks in MailingList, Forums and CVS and at this stage, the type of activities to be traced will be observing and establishing contact through reading and commenting for both participants. And in this instance, either a title in a discussion forum and MailingLists or an artifact thread can be taken as a process. The same applies for the remaining two stages to be considered mutatis mutandis.

### 3.2. Act-Trace Algorithm and Results

In this paper, we introduce a simple algorithm, called *Act-Trace*, for the purpose of locating, tracking and tracing related trace activities and their corresponding authors as well as the duration of activities execution.

Let $S_c$ denote the source code file;
Let $S_b$ denote the person having submitted the code file;
Let $S_r$ denote the reviewer and $R_c$ denote his/her comments
A $S_r$ can be either an *initiator* [ anybody submitting a review about a piece of code] or *responder* [ anybody reacting to a submitted comment]

**Act-Trace [Activity-Trace] Algorithm:**

***Input:***
Given a $S_c$

***Processing:***
Loop Through $S_c$
For every $R_c$
Return *initiator* ("Name", "Review Comment" ," Date"),
*responder* (*initiator* ("Name", "Review Comment" ," Date");
End Loop

***Output:***
TraceLog

In order to conduct our experiment, we chose OpenStack [1] as our data source. Our case study depicts a representation of conducted activities between members of an OSS group for a given period of time. We consider activities that occurred between the 3rd of May 2012 and 19th of July 2012 on reviews of submitted source-code.

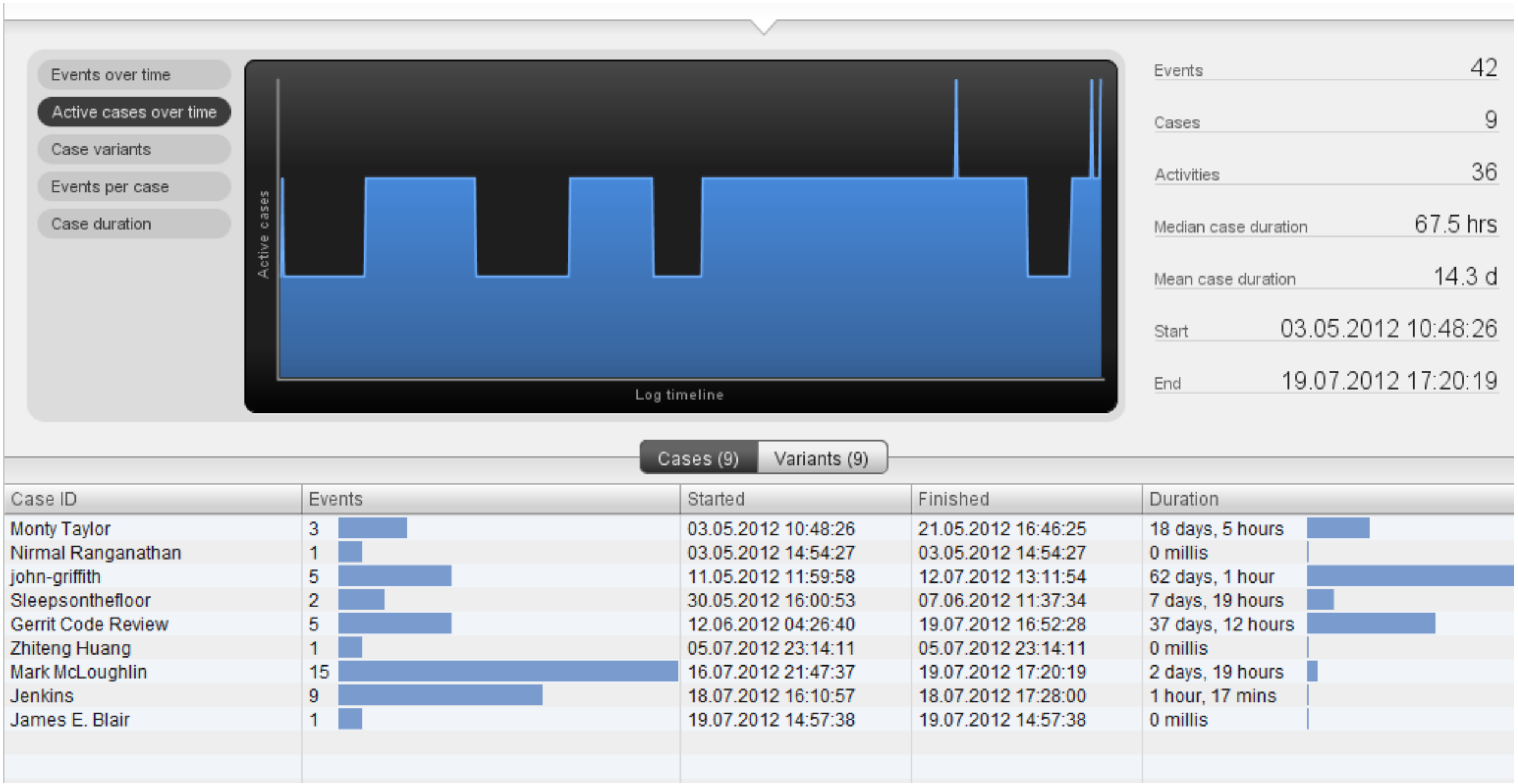

| Case ID | Events | Started | Finished | Duration |
|---|---|---|---|---|
| Monty Taylor | 3 | 03.05.2012 10:48:26 | 21.05.2012 16:46:25 | 18 days, 5 hours |
| Nirmal Ranganathan | 1 | 03.05.2012 14:54:27 | 03.05.2012 14:54:27 | 0 millis |
| john-griffith | 5 | 11.05.2012 11:59:58 | 12.07.2012 13:11:54 | 62 days, 1 hour |
| Sleepsonthefloor | 2 | 30.05.2012 16:00:53 | 07.06.2012 11:37:34 | 7 days, 19 hours |
| Gerrit Code Review | 5 | 12.06.2012 04:26:40 | 19.07.2012 16:52:28 | 37 days, 12 hours |
| Zhiteng Huang | 1 | 05.07.2012 23:14:11 | 05.07.2012 23:14:11 | 0 millis |
| Mark McLoughlin | 15 | 16.07.2012 21:47:37 | 19.07.2012 17:20:19 | 2 days, 19 hours |
| Jenkins | 9 | 18.07.2012 16:10:57 | 18.07.2012 17:28:00 | 1 hour, 17 mins |
| James E. Blair | 1 | 19.07.2012 14:57:38 | 19.07.2012 14:57:38 | 0 millis |

Figure 3: Statistical Information about Constructed Log

In Figure 3, we project general key information as obtained to understand the chunk of data we set to analyze. The figure depicts a histogram that represents the execution of activities as related to the duration or time taken to execute them. For a specified period of time, as specified on the right hand side of the graph, a total of nine individuals have at some extent reviews the same piece of code and rect to one another's comments. The most vocal individual happen to be Mark McLaughlin with a total of 15 messages or comments posted in reactions to what has been done or in response to comments made on his piece of code.

Figure 4 below on the other hand is the Process Model that graphical shows the flow of occurrence of all these activities. Each path of the workflow net represents a separate individual activities and the sequence in which they work as specified by the corresponding timestamp.

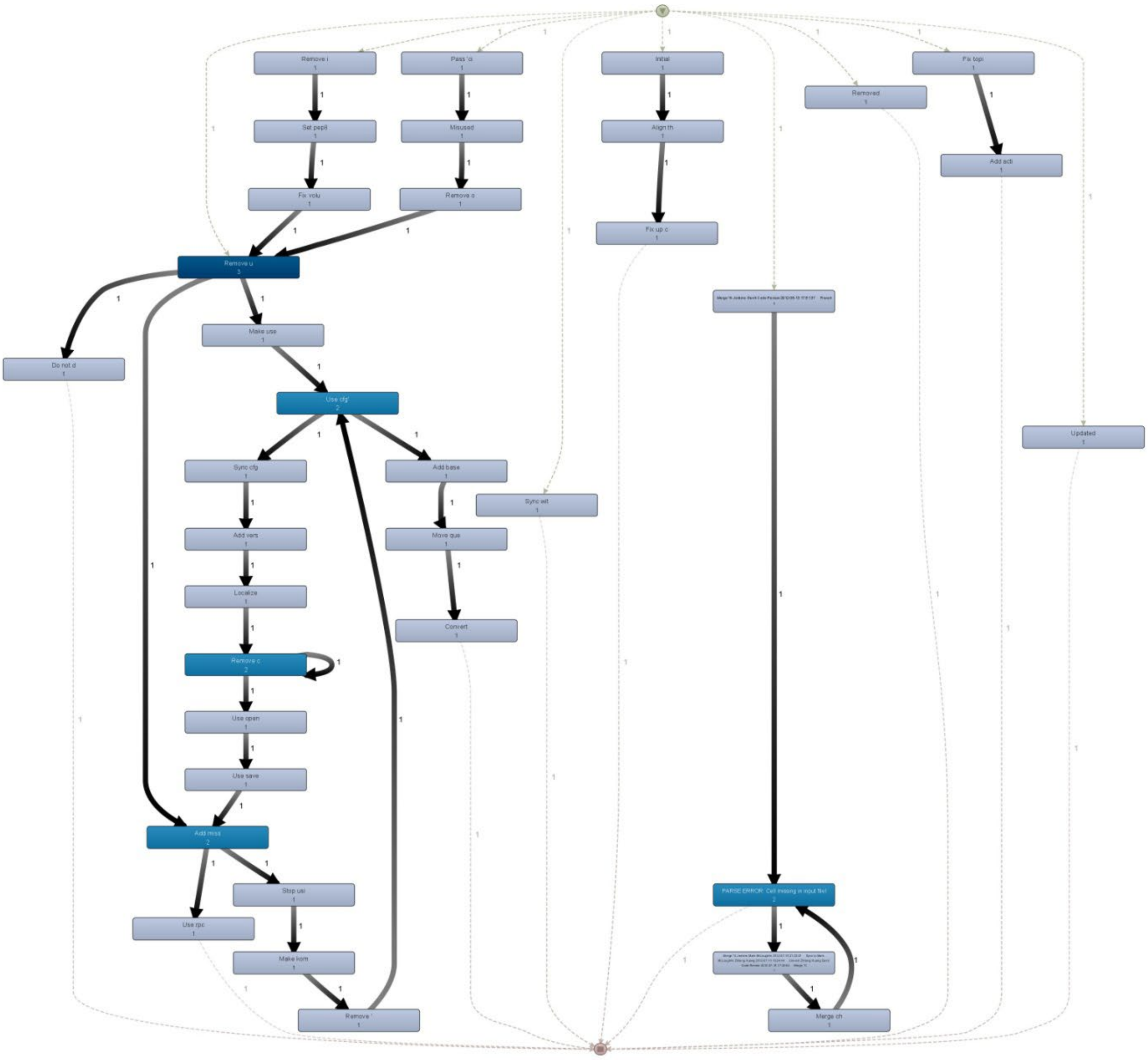

**Figure 4: Generated Process Model representing the flow of activities for activities as performed by each of the participants**

Given the constraints of space, we  choose just two of of the pathms and zoom on some activities in order to give the reader a better sens of what we are talking about:John and Gerrit, these depict they activities they worked on as a result from anyone [also shown] and the duration for the completion of the tasks as shown in figures 5 and 6.

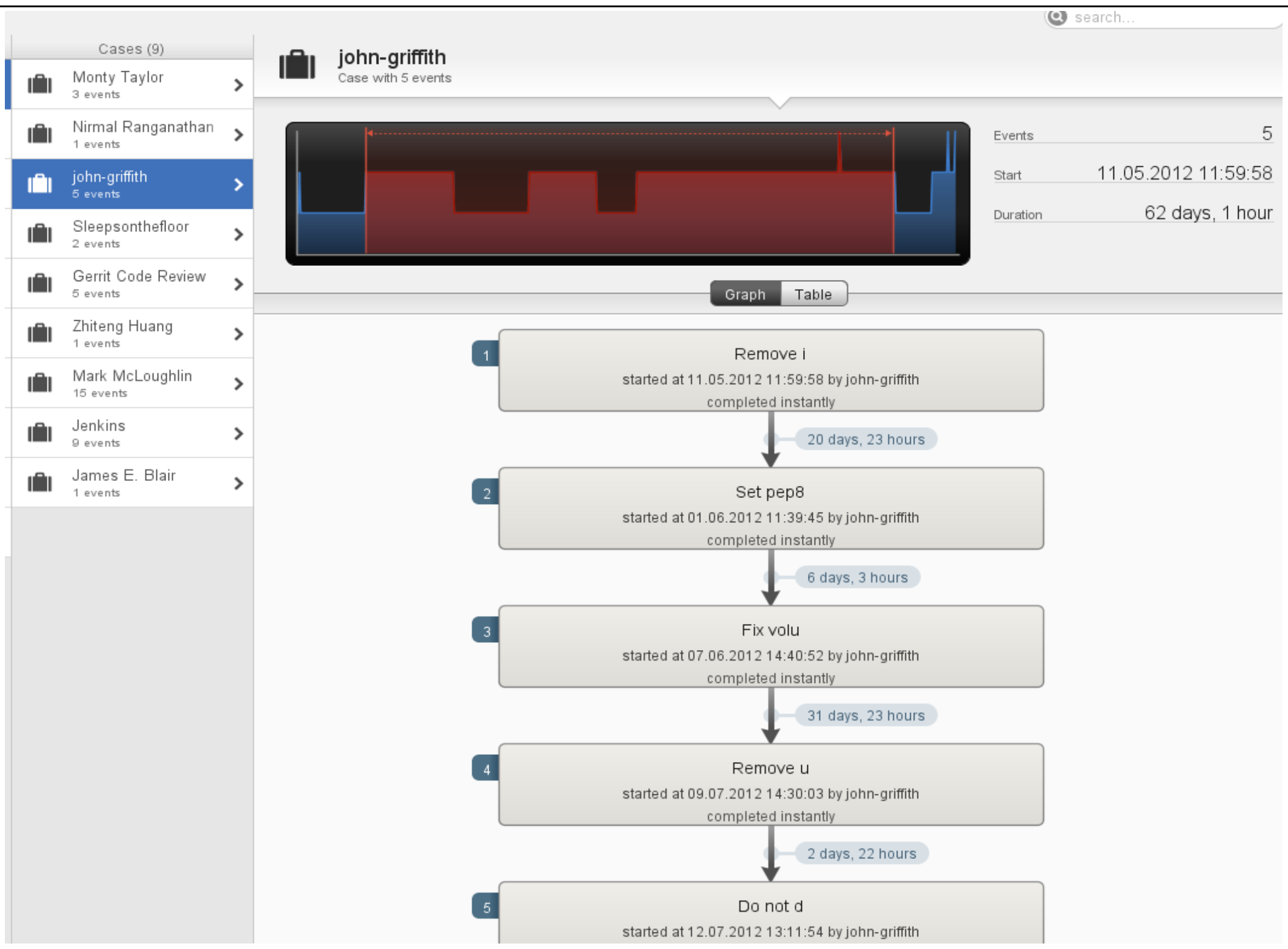

**Figure 5: A snapshot of John Griffin's sequence of occurrence for first five comments committed**

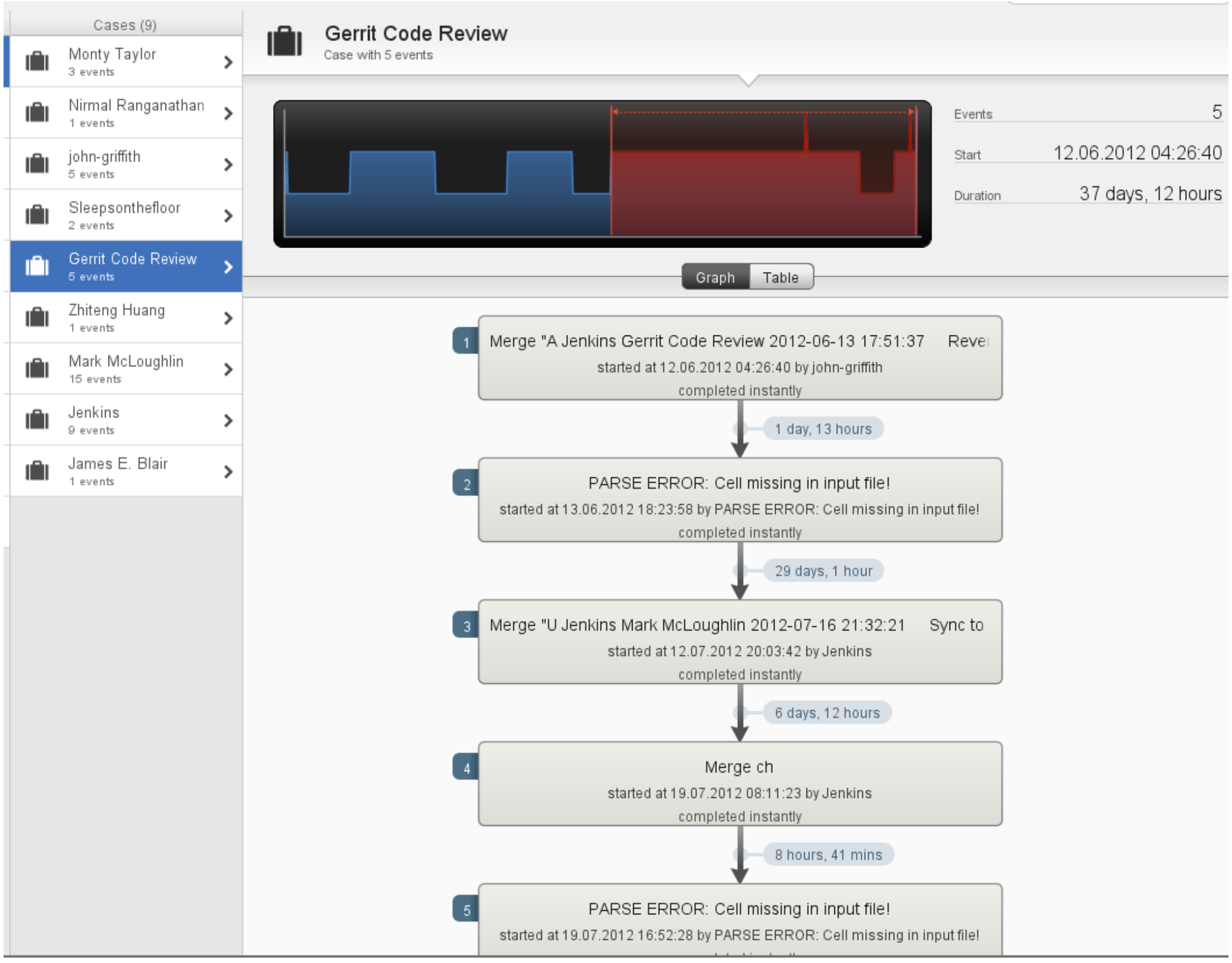

**Figure 6: A snapshot of Gerrit Code Review's sequence of occurrence for first five comments committed**

In order to express the sense of collaboration between these participants, we also opt to generate a social graph. This can give us an indiation of who reviewed whose source code and trace the relationship between them if needed as depicted in figure 7. When deriving such roles, the focus is on the relations among individuals (or groups of individuals) acting in the process. In this graph, each reviewer represents a node and the edge between reviewers is established when one has reacted to another's comment. Simply put, when a responder has posted a comment to the initiator's comment, that creates a link between the two.

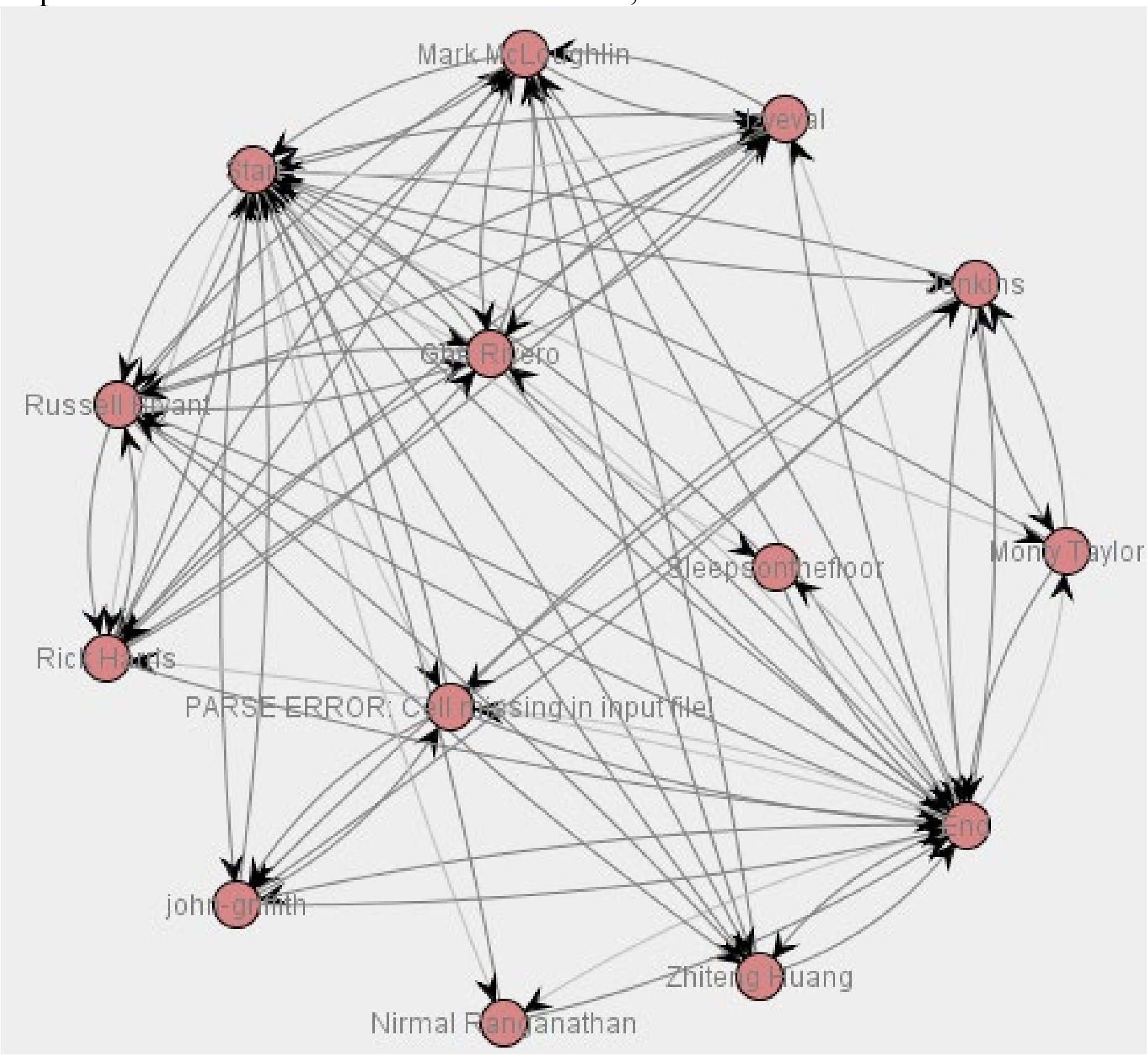

**Figure 7: A Social Network for Collaborating Reviewers**

## 4. CONCLUSION

Producing workflow logs and process mining them provides a ton of benefits given that the main first purpose that the method serves is to discover how people and procedures really work. An example could be the understanding of flow of patients in a hospital. While information about activities in these environments is available, the format in which usch information can be found in OSS enviornments is different and makes the process complex. In this paper, we introcude a simple algorithm that can trace and regroup related activities for reviewers of source code in an online community of development. The primary objective of this approach is to demonstrate the applicability of process mining in Open Source Environments as this is still very much lacking.

We made use of data from the OpenStack community and considered information about users over a defined period of time using the Act-Trace algorithm. The produced log was analysed using DISCo process mining tool [10].

**BIBLIOGRAPHY OF AUTHOR**

| | |
|---|---|
| 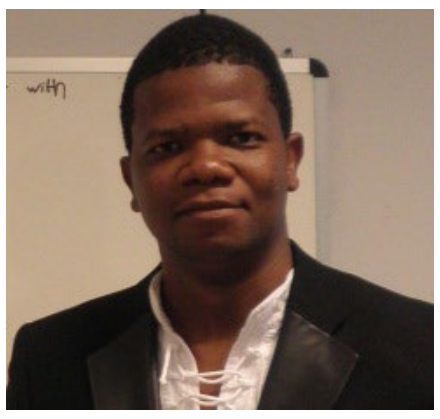 | Patrick Mukala is currently a PhD Candidate in Computer Science at the University of Pisa in Italy. His doctoral work includes process and data mining in FLOSS repositories. |